\documentclass{article}
\pdfoutput=1
\usepackage{arxiv}

\usepackage[utf8]{inputenc} 
\usepackage[T1]{fontenc}    
\usepackage{hyperref}       
\usepackage{url}            
\usepackage{booktabs}       
\usepackage{amsfonts}       
\usepackage{nicefrac}       
\usepackage{microtype}      
\usepackage{lipsum}

\usepackage{amsmath}
\usepackage{graphicx,psfrag,epsf}
\usepackage{enumerate}
\usepackage{natbib}

\usepackage{adjustbox}
\usepackage{multirow}
\usepackage{xcolor}
\usepackage{caption}
\usepackage{subcaption}
\usepackage{chngcntr}
\usepackage[flushleft]{threeparttable}

\renewcommand{\theequation}{\arabic{equation}}

\title{\bf CausalDeepCENT: Deep Learning for Causal Prediction of Individual Event Times}

\author{
Jong-Hyeon Jeong \thanks{Corresponding Author: jjeong@pitt.edu}\\
Department of Biostatistics\\
Graduate School of Public Health\\
University of Pittsburgh, Pittsburgh, USA\\
\and {\bf Yichen Jia}\\
Department of Biostatistics\\
Graduate School of Public Health\\
University of Pittsburgh, Pittsburgh, USA\\
}

\begin{document}
	\maketitle

\begin{abstract}
Deep learning (DL) has recently drawn much attention in image analysis, natural language process, and high-dimensional medical data analysis. Under the causal direct acyclic graph (DAG) interpretation, the input variables without incoming edges from parent nodes in the DL  architecture maybe assumed to be randomized and independent of each other. As in a regression setting, including the input variables in the DL algorithm would reduce the bias from the potential confounders. However, failing to include a potential latent causal structure among the input variables affecting both treatment assignment and the output variable could be additional significant source of bias. The primary goal of this study is to develop new DL algorithms to estimate causal individual event times for time-to-event data, equivalently to estimate the causal time-to-event distribution with or without right censoring, accounting for the potential latent structure among the input variables. Once the causal individual event times are estimated, it would be straightforward to estimate the causal average treatment effects as the differences in the averages of the estimated causal individual event times. A connection is made between the proposed method and the targeted maximum likelihood estimation (TMLE). Simulation studies are performed to assess improvement in prediction abilities of the proposed methods by using the mean square error (MSE)-based method and rank-based $C$-Index metric. The simulation results indicate that improvement on the prediction accuracy could be substantial particularly when there is a collider among the input variables. The proposed method is illustrated with a publicly available and influential breast cancer data set. The proposed method has been implemented by using PyTorch and uploaded at \verb|https://github.com/yicjia/CausalDeepCENT|.
\end{abstract}
\noindent {\it Keywords:}  Causal Inference; Neural Network; Propensity Scores; Survival Analysis; Time to Event

\section{Introduction}
\label{s:intro}

In many research areas such as biology, economics, engineering, medicine, politics, and sociology, time-to-event data are frequently encountered. Particularly in medical studies, investigators might be interested in time to liver failure from diagnosis of a liver disease \citep{vanwagner2018} or time to a breast cancer recurrence or death after a surgery \citep{fisher2002,curtis2012genomic} from randomized or observational studies. Time-to-event outcome is often dichotomized to be simplified, but such dichotomization results in a substantial loss of information. An important feature of time-to-event data is right-censoring, which occurs when a subject does not experience an event of interest by the end of the study period. The censored observations prohibit observation of the events of interest but still preserves partial information that the true event time would lie beyond the observed censoring time point.  Time-to-event data as exemplified above can be summarized by using the well-known hazard function, survival function, or the mean or quantile survival time, but investigators may also want to know a single valued time to an event of interest with a certain level of confidence to arrange an intervention strategy at an individual level. For example, physicians might be interested in predicting an individual time to a breast cancer recurrence or death given the patient’s characteristics and genetic signatures and identifying an optimal intervention to delay his/her time to a failure by a certain number of years. In general, it is challenging to predict an individual event time using the traditional statistical approaches for censored time-to-event data due to poor prediction accuracy, although the estimates of the population characteristics such as survival probabilities and model-based prognostic effects would be still reliable \citep{Henderson2005}. Without prognostic factors, a plausible approach to estimate an individual event time might be by inverting the nonparametric Kaplan-Meier survival curve estimates for a specific quantile with an associated prediction interval. However, the traditional model-based approaches often depend on the log-linear or parametric assumptions between the censored outcome and the prognostic factors, so it would not be robust to a model misspecification, introducing further variability in individual prediction. In addition, to use the predicted values for an intervention at an individual level, it would be crucial to accurately estimate counterfactual outcome without the intervention.

Deep learning (DL) has recently drawn much attention in image analysis, natural language process, and high-dimensional medical data analysis. In the field of statistics, it has been also shown that the DL  algorithm could predict nonlinear patterns more accurately compared to traditional linear quantile regression and nonparametric quantile regression \citep{jia2022a}. With observational medical data, data-driven machine learning algorithms such as DL  and random forests are often adopted to predict patients’ outcomes given high dimensional input variables. However, the use of machine-learning approaches based on the observed data to model causality for intervention effects might result in unjustifiable and biased consequences due to potential confounding and selection biases \citep{prosperi2020}. Deep learning has an inherent causal DAG \citep{pearl1995} interpretation with incoming directed edges from the previous layers. The input variables into the DL  algorithm can be viewed as the independent predictors as in a regression model, so that the effect of each input variable on the output variable can be interpreted as the total causal effect going through all possible paths between the input variable and the output variable, conditioning on the other input variables. Under the DAG interpretation of the DL  algorithm, furthermore, without incoming edges or parent nodes, the input variables are assumed to be randomized and independent of each other. In observational data, however, a combination of the input variables might also affect the treatment assignment inducing an imbalance across the intervention groups.

\subsection{Motivation from a real data set}

\begin{figure}
    \centering
    \includegraphics[width = 0.7\linewidth,keepaspectratio]{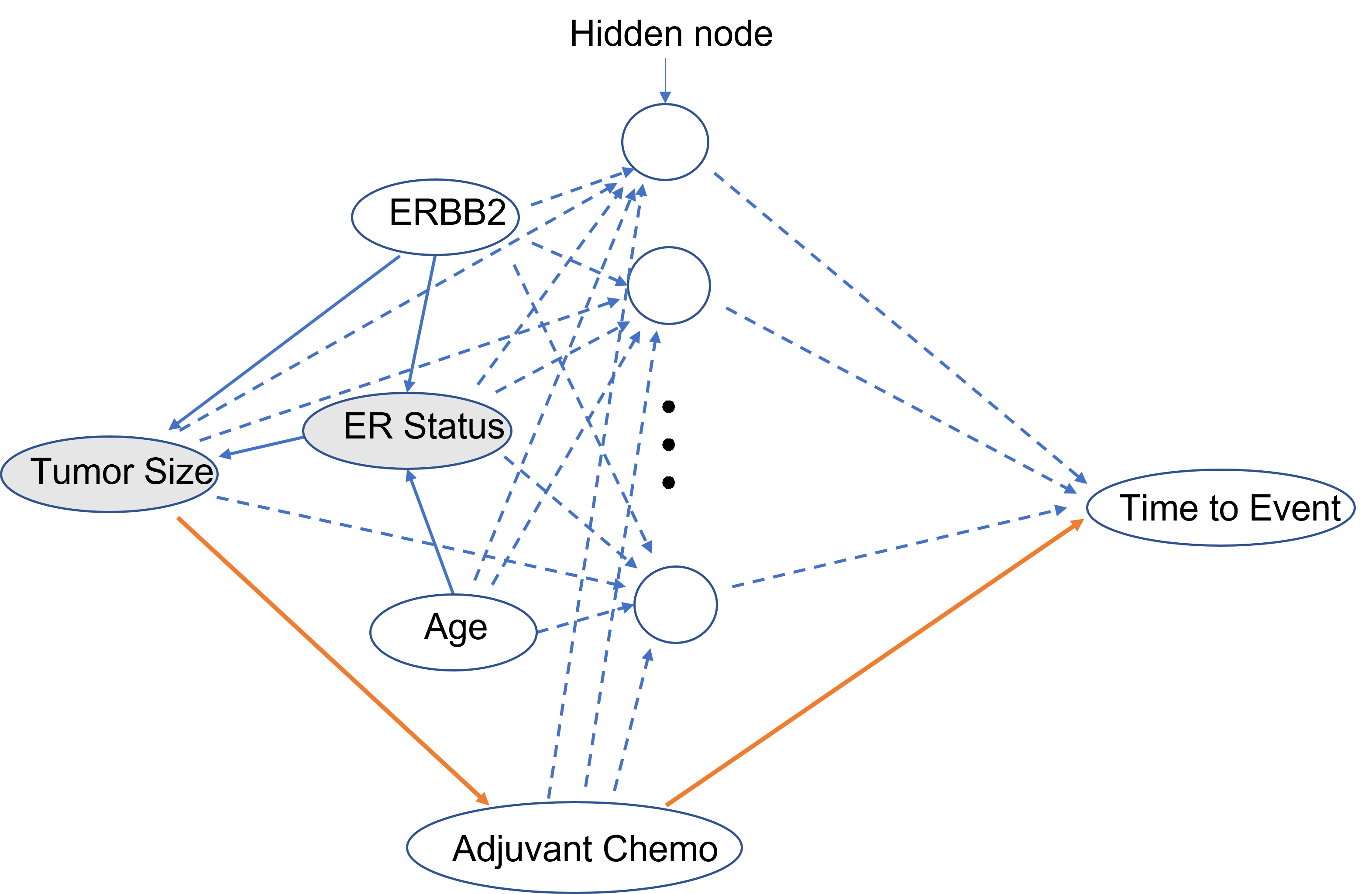}
    \caption{Potential bias due to causal structure among a subset of the input variables in METABRIC data set.}
    \label{fig:bias}
\end{figure}

We begin with a motivating example of a real data set. The METABRIC (Molecular Taxonomy of Breast Cancer International Consortium) data set \citep{curtis2012genomic} is a collection of over 2,000 clinically annotated primary fresh-frozen breast cancer specimens from tumour banks in the UK and Canada, together with a collection of clinical variables. Nearly all estrogen receptor (ER)-positive and/or lymph node (LN)-negative patients did not receive chemotherapy, whereas ER-negative and LN-positive patients did. Additionally, none of the HER2-positive patients received trastuzumab (hornomal therapy). The primary outcome was time to death subject to censoring. To model this type of observational data, a combination of clinical variables and gene signatures are typically included in the model as the predictors after an initial screening procedure. In our data analysis to be presented in Section 5, we have included 4 genes via their expression levels (MKI67, EGFR, PGR, and ERBB2) with 10 clinical features (estrogen receptor (ER) status, human epidermal growth factor receptor 2 (HER2) status, progesterone receptor (PR) status, tumor size, age at diagnosis, number of positive lymph nodes, menopausal status, Nottingham prognostic index, hormonal therapy status, and radiotherapy status), all of which individually had significant association with the primary outcome, i.e., time to death. Using a subset of the predictor list above, Figure \ref{fig:bias} demonstrates potential yet plausible causal relationship among the well-known predictors for breast cancer, which can affect both the primary outcome and treatment assignment, an adjuvant chemotherapy. Specifically, the expression level of the gene ERBB2 as a hormonal growth factor can affect tumor size \citep{gilcrease2009} and ER status \citep{pinhel2012}, and age can be associated with the ER status \citep{benz2008}, among other things, which can also affect tumor size. Prescription of an adjuvant chemotherapy can be affected by tumor size, which is expected to have an impact on the primary outcome. In this diagram, note that tumor size and ER status could be colliders, and once they are included as the input variables in the model, they could open all the non-causal paths that go through them, introducing bias. The dotted blue lines indicate the paths the standard deep leaning algorithm would take for the optimization procedure, ignoring the causal relationship among the input variables indicated by the solid lines.

\subsection{Related work and contribution}

Causal prediction through DL  algorithms has been recently drawing much attention. \cite{koch2021} reviewed four different approaches to deep causal estimation including (i) meta-learners \citep{binyu2019} like S-learners, T-leaners, and X-learners, (ii) balancing through representation learning such as TARNet \citep{Shalit2017} and CFRNet \citep{johansson2018,johansson2020}, (iii) extension with inverse propensity score weighting such as targeted maximum likelihood estimation (TMLE, \cite{vanderlaan2011}) and Dragonnet and targeted regularization \citep{shi2019}, and (iv) adversarial training of generative models such as GANITE \citep{yoon2018}. In particular, TARNet utilizes the loss functions that minimize the mean square errors (MSE) of the observed vs. predicted causal outcomes after representation learning, i.e., after deconfounding the treatment from the outcome by forcing the treated and control covariate distributions to get closer together, and CFRNet additionally minimizes the distance between treatment indicator and covariate distributions. Dragonnet minimizes the loss function that consists of the MSE of the predicted outcome and the distance between the treatment indicator and the propensity score, based on the covariates decorrelated from the treatment. TMLE minimizes a specific efficient influence curve (EIC) after plugging in the predicted outcome and estimated propensity scores, which provide a doubly robust estimator for the average treatment effect (ATE). Another related reference to the proposed work is DAGSurv \citep{sharma2021} where the causal DAG is estimated from the given data set and used as an input into the DL  algorithm to predict survival probability. In the DL  framework under noncausality, \cite{baek2021survival} integrated a pre-trained hazard ratio network (DeepSurv) and a newly proposed distribution function network to predict individual event times, \cite{jing2019deep} utilized a loss function that combined an extended mean squared error (MSE) loss and a pairwise MSE ranking loss to predict individual event times, and \cite{jia2022b} proposed to predict individual individual times by introducing a loss function for the DL algorithm that combines the mean square error and the concordance index, including a novel extension to competing risks data.

The algorithm proposed in this study can be viewed as a one-step variation of the TMLE approach by minimizing the weighted least residual sum of squares of the predicted individual event time (possibly censored) via the inverse propensity score weighting. The resulting predicted individual event times, or its distribution, will have a causal interpretation, and hence will be referred to as {\em estimated causal individual event times} or {\em estimated causal event time distribution}. It is noted that once the causal event time distribution is estimated, the causal average treatment effects (for the treated) can be also readily calculated at the population level for the censored time-to-event data as the mean difference in the estimated causal individual event times between the intervention groups.

In Section 2, we review some preliminaries including the inverse propensity score weighting and potential outcomes framework for causal inference on time-to-event data. In Section 3, we present the weighted causal DL  with weighted loss function to predict individual event times with or without censoring. A connection is also made between our proposed work and the TMLE method. Simulation studies are presented in Section 4 when there is an imbalance between the intervention groups induced by an association among the input variables that affects treatment assignment and the outcome variable. The proposed algorithm is illustrated with a publicly available and influential breast cancer data set in Section 5. Finally, we conclude with a brief discussion in Section 6.

\section{Preliminaries}

In this section, we review the propensity score weighting method and introduce the potential outcomes framework for censored survival data for causal inference. Throughout the paper, the terms of input variables, covariates, confounding factors, and predictors, will be used interchangeably.

\subsection{Propensity Score Weighting}
The propensity score is a subject's probability of receiving treatment, which is particularly useful when exact matches for each case are not feasible possibly due to high dimensional data. The propensity scores can be used in various ways to balance the potential confounding factors, but in this study the propensity score weighting method will be adopted. This approach uses the inverse probability of treatment weights (IPTW) to make the treatment and the control groups balanced on the propensity scores and therefore also likely on the individual covariates. There are two popular types of the propensity score weights; the ATE type and the ATT type. Suppose $\pi_i$ denotes the propensity score, i.e., the probability that each subject would be assigned to the treatment group given a vector of covariates. First, the ATE weights are defined as
\begin{equation}
    w_i^{(ATE)}=\frac{Z_i}{\pi_i}+\frac{1-Z_i}{1-\pi_i}, \label{eqn;1}
\end{equation}
implying that a subject in the control group would contribute more counterfactual information toward the treatment group if his/her propensity score is higher and a subject in the treatment group would contribute more counterfactual information toward the control group if his/her propensity score is lower. This allows for estimation of the average causal effect of switching treatment. Second, the ATT weights are defined as
\begin{equation}
    w_i^{(ATT)}=Z_i+\frac{\pi_i(1-Z_i)}{1-\pi_i}, \label{eqn;2}
\end{equation}
which assigns the weight of 1 to all the observations in the treatment group and more weights to the observations with higher propensity scores in the control group, extracting more counterfactual information from the control group toward the treatment group. This allows for estimation of the average causal effect of taking away the treatment.
For causal inference, the weights can be incorporated into a weighted regression setting to estimate causal treatment effects such as the ATE or ATT at a population level. In this study, the propensity scores will be calculated by using the treatment variable as the outcome variable and the other input variables as covariates in a logistic regression. Then the propensity scores will be transformed into the weights in equations (\ref{eqn;1}) or (\ref{eqn;2}), which will be included in the loss function of the DL  algorithm.

\subsection{Potential Outcomes Framework for Censored Time-to-Event Data}

For a subject $i$, $Y_i(0)$ and $Y_i(1)$ denote the {\em counterfactual} individual event times for the control ($Z_i=0$) and treatment ($Z_i=1$) groups, respectively. Once the treatment assignment is realized, the {\em observable} individual event time is determined by $Y_i=(1-Z_i)Y_i(0)+Z_iY_i(1)$. Under the consistency and ignorability assumptions \citep{rubin1974}, the causal average effect $E\{Y(1)-Y(0)\}$ can be estimated based on the observed outcomes as $E(Y|Z=1)-E(Y|Z=0)$. When there is an independent right censoring, $C_i(0)$ and $C_i(1)$ denote the {\em counterfactual} outcome of individual censoring times for the control and treatment groups, respectively.  The {\it observable} individual censored time is denoted by $C_i=(1-Z_i)C_i(0)+Z_iC_i(1)$ and finally the {\it observed} survival time is determined as $T_i=\min(Y_i,C_i)$  with an event indicator $\delta_i=I(Y_i<C_i)$.

\section{Weighted Causal Deep Learning}
\subsection{Model Architecture}
\begin{figure}
    \centering
    \includegraphics[width = 0.7\linewidth,keepaspectratio]{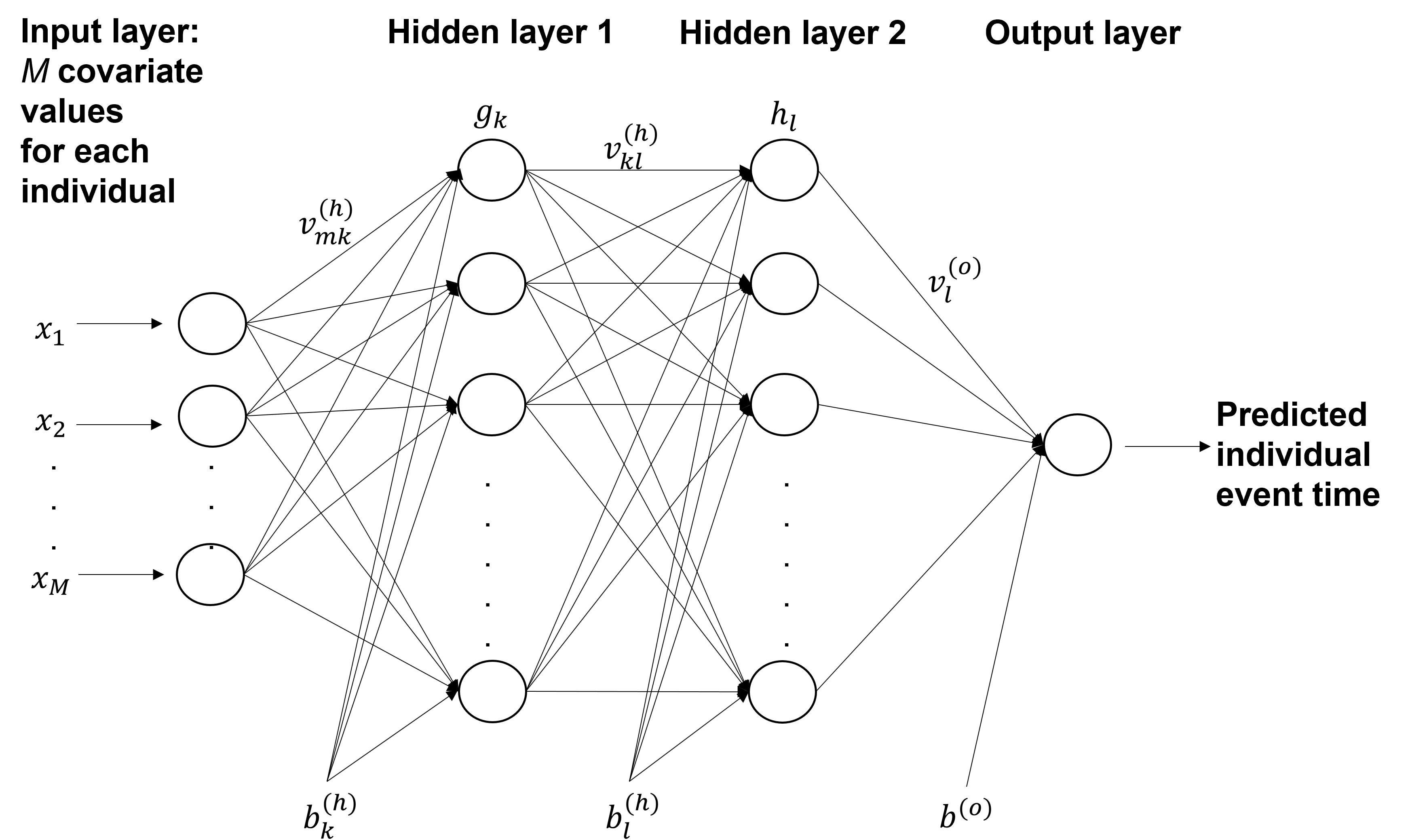}
    \caption{Fully-connected DL  architecture.}
    \label{fig:archi}
\end{figure}

We propose a deep feed-forward neural network to predict {\em individual event time}. Figure \ref{fig:archi}  shows the basic model architecture. The hidden layers of the network are dense, i.e. fully connected by the nodes. The output of a hidden layer is given by applying the activation function to the inner product between the input and the hidden-layer weights plus the hidden-layer bias. For example, suppose there are $M$ input variables and two hidden layers. Then, the output of the $k^{th}$ hidden node for the first hidden layer would  be
$$g_k = f_1\left(\sum_{m=1}^{M} x_{m} \nu_{mk}^{(h)}+b_k^{(h)}\right),\quad k=1,2,...,K,$$
and the output of the $l^{th}$ hidden node in the second hidden layer would be
$$h_l = f_2\left(\sum_{k=1}^K g_k \nu_{kl}^{(h)}+b_l^{(h)}\right),\quad l=1,2,...,L,$$
where $f_1(\cdot)$ and $f_2(\cdot)$ denotes the activation functions for the hidden layers, and $\nu^{(h)}$ and $b^{(h)}$ represent the hidden-layer weights and bias, respectively, both of which get updated at each training iteration. Note that a single layer neural network with linear activation function could approximate the ordinary linear regression without censoring. The bias terms allow for shifting the activation function outputs toward the left or right \citep{gallant1993neural, bishop1995neural,reed1999neural}.
Lastly, the output layer of the network is a single node with a linear activation function which gives the estimate of the conditional individual event time $\hat{T}_i$ on a log-scale for the $i^{th}$ subject as
$$\log{\hat{T}_i} = \sum_{l=1}^L h_{l,i} \nu_{l}^{(o)}+b^{(o)},$$
where $\nu^{(o)}$ and $b^{(o)}$ denote the output-layer weights and bias, respectively. \cite{vidal2017} provided an extensive literature review on several mathematical properties of deep networks such as global optimality, geometric stability, and invariance of the learned representations. \cite{rob2021} showed that the optimization of the hidden multi-layer weights can be viewed as a cubic regression problem.

\begin{figure}
    \centering
    \includegraphics[width = 0.9\linewidth,keepaspectratio]{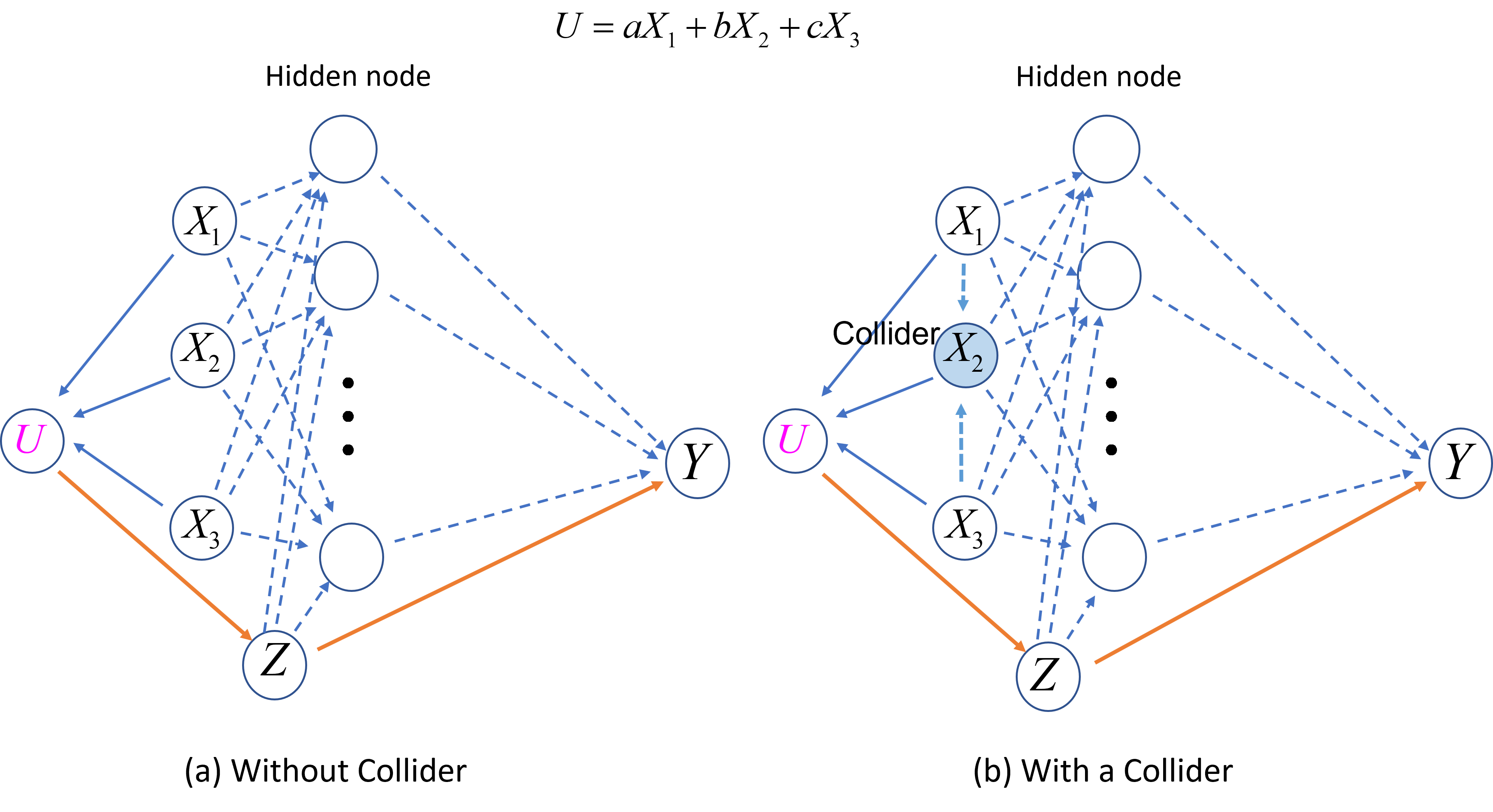}
    \caption{Examples of DAG representation of the DL  architecture.}
    \label{fig:dag}
\end{figure}

This standard DL  architecture will be considered as the generic causal diagram that implies all the input variables are randomized and independent of each other, but our true causal DL  architecture where our data are generated from will assume a hidden dependent structure such as the treatment variable being affected by a linear combination of all the other input variables with or without colliders among the input variables. Figure \ref{fig:dag}(a) illustrates a simple example where there exists 4 input variables ($X_1$, $X_2$, and $X_3$ and a treatment assignment variable $Z$) as predictors, and a latent variable ($U$) as a linear combination of $X_1$, $X_2$, and $X_3$, which directly affects $Z$, possibly introducing imbalance in treatment assignment. The observable outcome $Y$ also has direct association with $X_1$, $X_2$, $X_3$, and $Z$ through the DL  algorithm. Under this setting, our conjecture is that the DL  algorithm would only optimize direct causal relations (dotted lines) between the input and output variables based on the observed data ignoring the causal paths going through the unobservable $U$, which might introduce bias in causal estimation. A larger bias might be introduced when there is a hidden collider effect among the input variables as in Figure \ref{fig:dag}(b), which opens more causal paths between the input and output variables that might be ignored by the standard deep leaning algorithm optimization procedure. If the variable $U$ is in fact observed and included as one of the input variables, then it becomes another collider that might introduce further bias. These scenarios will be considered in our simulation studies in Section 4.

\subsection{Loss Function}
In this section, we will develop a causal DL algorithm that utilizes (i) a new loss function including the propensity score weights to inherently adjust for the bias due to a potential latent causal structure among the input variables and (ii) a penalty term against the predicted individual event time being shorter than an observed censoring time when the censoring occurs first.

\subsubsection{No censoring (NC) cases}
The DL algorithms require a loss function to be minimized usually by the back propagation with gradient decent algorithm \citep{rumelhart1986learning}. When there is no censoring, the time-to-event outcome becomes an ordinary continuous variable, so the simple mean square error (MSE) can be used as the loss function. Mimicking the weighted linear regression setting commonly used to estimate the average causal treatment effects at the population level, we propose to include the level of counterfactual contribution of each observation directly into the DL loss function as the weights to predict individual event times.  For the uncensored cases, therefore, we propose the loss function as
\begin{equation}
    L_{NC}=\sum_{i=1}^n \hat{w}_i (\hat{T}_i-T_i)^2, \label{eqn;3}
\end{equation}
where $\hat{w}$  is the estimated ATE or ATT weights, $T$  is the observed event time, and $\hat{T}$  is the predicted event time. Therefore, the predicted individual event times from the DL algorithm that minimize the loss function $L_{NC}$  would have causal interpretation as adjusting for the bias due to the potential imbalance possibly caused by a latent causal structure among the input variables.

The loss function in (\ref{eqn;3}) can be interpreted in view of the TMLE method. It is well known from the semiparametric theory for the TMLE (Shi et al., 2019) that the solution to the following efficient influence curve (EIC) provides asymptotically unbiased and efficient estimates of the average treatment effects. For example, with the ATE weights, the EIC would have a form
\begin{equation}
    \frac{1}{n}\sum_{i=1}^n \left[\left\{\frac{Z_i}{\pi_i(x_i,Z_i=1)}-\frac{1-Z_i}{\pi_i(x_i,Z_i=0)}\right\}\left\{T_i-g(x_i,Z_i)\right\}+\left\{g(x_i,Z_i=1)-g(x_i,Z_i=0)\right\}\right]-ATE,\label{eqn;3.1}
\end{equation}
where $g(x_i,Z_i)$ is the underlying truth of the outcome variable, $\pi(x_i,Z_i)$ is the true propensity score function given the treatment variable $Z_i$ and a covariate vector $x_i$, and $\epsilon_i=T_i-g(x_i,Z_i)$ is the residual function for the entire data. This score-type function is also known to be Neyman orthogonal \citep{cher2017, neyman1959}, implying that inference on the ATE would be robust to small mistakes in nuisance parameters such as the predicted outcomes and estimated propensity scores, and hence sufficiently insensitive to replacement of $g(x_i,Z_i=z_i)$ and $\pi(x_i,Z_i=z_i)$ with their estimates $\hat{T}_i(Z_i=z_i)$ and $\hat{\pi}(x_i,Z_i=z_i)$. By setting the equation (\ref{eqn;3.1}) to 0, a plug-in doubly robust estimator for the ATE is given by
\begin{equation}
    \widehat{ATE}=\frac{1}{n}\sum_{i=1}^n \left[\left\{\frac{Z_i}{\hat{\pi}_i(x_i,Z_i=1)}-\frac{1-Z_i}{\hat{\pi}_i(x_i,Z_i=0)}\right\}\epsilon_i+\left\{\hat{T_i}(Z_i=1)-\hat{T_i}(Z_i=0)\right\}\right].\label{eqn;3.2}
\end{equation}
To calculate the ATE, the TMLE is a two-step procedure, the first step being an input-output modeling without adjusting for the inverse propensity score weights and the second step being the estimation step for the fluctuation parameter $\epsilon_i$ as a constant regression coefficient for the slope between the inverse propensity score weights and the observed outcome.It was noted \citep{koch2021} that this would be equivalent to minimizing the weighted term involving $\epsilon_i$ in equation (\ref{eqn;3.2}).

An alternative simplistic approach, therefore, would be to directly estimate $g(x_i,Z_i)$ that minimizes the optimization function
\begin{equation}
    \frac{1}{n}\sum_{i=1}^n \left\{\frac{Z_i}{\hat{\pi}_i(x_i,Z_i=1)}-\frac{1-Z_i}{\hat{\pi}_i(x_i,Z_i=0)}\right\}\left[T_i-g(x_i,Z_i)\right]^2,\label{eqn;3.3}
\end{equation}
which is equivalent to (\ref{eqn;3}). This implies that the proposed loss function is a $L_2$-norm optimization procedure and is expected to analytically provide unbiased estimate of the distribution of $\hat{g}(x_i,Z_i)=\hat{T_i}(Z_i)$ in terms of the population averages due to all the nice properties of the weighted least squares estimators. Once $g(x_i,Z_i=z_i)$ is estimated as the causal individual event times without censoring, the causal ATE can be estimated as
$$ATE=\frac{1}{n}\sum_{i=1}^n\left\{\hat{T_i}(Z_i=1)-\hat{T_i}(Z_i=0)\right\}.$$
The main goal of this paper is to introduce the loss functions weighted by the inverse propensity scores directly into the DL  algorithm to estimate individual event times, and hence the causal distribution of time-to-event data with or without censoring, which can be further utilized to estimate the average treatment effects.

\subsubsection{Independent right censoring (RC) cases}
The partial information from a right-censored observation that is typically incorporated into survival analysis is that the event of interest did not occur before the observed censoring time, contributing to the risk set up to that point. For the right censoring case, borrowing the idea of the loss function for censored data from \citet{jing2019deep}, we will propose a weighted MSE-based loss function for causal prediction as
\begin{equation}
    L_{RC}=\sum_{i=1}^n \hat{w}_i\left[I(\delta_i=1)(\hat{T}_i-T_i)^2+\lambda I(\delta_i=0)I(\hat{T}_i < T_i)(\hat{T}_i - T_i)^2\right].\label{eqn;4}
\end{equation}
This loss function includes the estimated weights ($\hat{w}$), the squared distance between observed event time ($T$ ) and predicted event time ($\hat{T}$) when the true event time is observed ($I(\delta=1$)), a penalty term if the predicted event time is shorter than the observed censoring time when the event time is censored ($I(\delta=0$)), and a hyperparameter $\lambda$  to be tuned. The DL  algorithm using the loss function $L_{RC}$ will be an extension of the TMLE to the right-censored data and provide estimated causal individual event times while ensuring that the predicted event times be greater than the censoring time and adjusting for an imbalance caused by the latent causal structure among the input variables.

\subsection{Optimization}
The optimization of the neural network is achieved by the gradient descent process in the backpropagation phase \citep{rumelhart1986learning} to minimize the loss function. The minimum can be obtained as the derivative of the loss function with respect to the hidden-layer weights through the chain rule involving the inputs and outputs of the activation functions reaches 0. The gradient descent method is often used to iteratively search for the minimum of the loss function by updating old hidden-layer weight values by the amount of the slope of the loss function at those values multiplied by a learning parameter (usually small) that controls the size of convergence steps, since it is often impossible to obtain a closed form solution for the hidden-layer weights including the biases in the DL  algorithm. The backpropagation algorithm through the gradient descent method was shown to be simple and computationally efficient \citep{Goodfellow-et-al-2016}. Optimizers implemented in PyTorch typically include Stochastic Gradient Descent (SGD), Adaptive gradient optimizer (AdaGrad) \citep{duchi2011adaptive}, AdaDelta \citep{zeiler2012adadelta}, Adam (adaptive moment estimation) \citep{kingma2014adam}, and Adamax (a variant of Adam based on the infinity norm), among other things. We used the optimizer Adam in this paper.

\subsection{Hyperparameter Tuning}
Hyperparameter tuning is essential in machine learning methods such as neural network and random forests since each different training run of the algorithm could provide a different output even with the same set of hyperparameters. In addition, a model with more layers and more nodes per layer tends to capture the nonlinear patterns in the data better, so hyperparameter tuning is also crucial to prevent overfitting. In this paper, hyperparameters involved in neural network, including number of hidden layers (1, 2, and 3 layers), number of nodes in each layer (32, 64, 128, and 256 nodes), learning rate (0.01 and 0.001), dropout rate (0.1, 0.2, 0.3, and 0.4), and the tuning parameter $\lambda$ (0, 0.01, 0.1, 1, 10), were tuned in the training data sets by using the Ray Tune package implemented in PyTorch. Since it is well known that the overly inflated number of epochs, defined as the number times that the learning algorithm will work through the entire training data set, can cause overfitting, an early stopping algorithm  was adopted to monitor the error loss in the test data set during the parameter tuning procedure.

\subsection{Model Evaluation}
\label{metric}

To evaluate the prediction performance in this paper, we use the concordance-index ($C$-index) \citep{harrell1984regression} and the mean squared error (MSE). For the simulated data with or without censoring, the MSE will compare true and predicted values to assess the model accuracy. For real data analysis, however, since the true event times are not observed for the censored observations, the MSEs are only calculated over event times.

The $C$-index is one of the most common metrics used to assess the prediction accuracy of a model in survival analysis, which is a generalization of the area under the ROC curve (AUC) that takes censoring into account \citep{heagerty2005survival}. The $C$-index is also related to the rank correlation between the observed and predicted outcomes. Specifically, it is the proportion of all comparable pairs that the predictions are concordant. For example, two samples $i$ and $j$ are comparable if there is an ordering between the two possibly censored outcomes as $T_i<T_j \text{ and } \delta_i=1$. Then a comparable pair is concordant if a subject who fails at an earlier time point is predicted with a worse outcome, which is the predicted individual event time in our case, i.e. $\hat{T}_i < \hat{T}_j$. The value of $C$-index is between 0 and 1 where 0.5 indicates a random prediction and 1 is a perfect association between predicted and observed outcomes.

It is worth noting that the rank-based methods like the $C$-index are not sensitive to small differences in discriminating between two models \citep{harrell1996multivariable}. For instance, the $C$-index considers the (prediction, outcome) pairs (0.01, 0), (0.9, 1) as no more concordant than the pairs (0.05, 0), (0.8, 1). In other words, the $C$-index focuses on the order of the predictions instead of the actual deviations of the prediction from the observed outcome such as the MSE.

\subsection{Prediction Uncertainty}
\label{pi}
Understanding the uncertainty of a predicted outcome is crucial in practice. It is worth noting that prediction interval is different from confidence interval in that the latter quantifies the uncertainty in an estimated population parameter while the former measures the uncertainty in a single predicted outcome. Several methods for obtaining the prediction interval have been proposed for the DL  algorithm \citep{de1998prediction, nix1994estimating, heskes1997practical}. In this paper, we utilize the dropout method as Bayesian approximation of the Gaussian process to obtain model uncertainty \citep{gal2016dropout}. Dropout was first proposed as a regularization method to avoid over-fitting in neural network by randomly dropping out a portion of nodes in a given layer \citep{hinton2012improving, srivastava2014dropout}. Specifically, suppose that we have the inputs $\mathbf{X}=[\mathbf{x}_1,\mathbf{x}_2,...,\mathbf{x}_n]^T$, the outputs $\mathbf{Y}=[\mathbf{y}_1,\mathbf{y}_2,...,\mathbf{y}_n]^T$, and $\mathbf{\nu}=[\mathbf{V}_1,\mathbf{V}_2,...,\mathbf{V}_L]^T$, where $L$ is the number of layers and $\mathbf{V}_i$ is a weight matrix connecting $(i-1)^{th}$ and $i^{th}$ layers. The dropout procedure is to first generate binary vectors from a Bernoulli distribution with a probability of success, and then multiply them to the weight matrices, dropping out the columns of the weight matrices randomly. The binary vectors are sampled for every layer input point and every forward pass through the model, and the same values are used in the back propagation phase. Dropout is typically used during the training stage but, using Monte Carlo Dropout \citep{gal2016dropout}, i.e. enabling dropout at the test stage and repeating the prediction multiple times, can also provide the model uncertainty.  Assuming that the model depends only on $\nu$ as the sufficient statistics, \cite{gal2016dropout} showed that the predictive distribution for a new input point $\mathbf{x}^*$ is given by
$$p(\mathbf{y}^*|\mathbf{x}^*,\mathbf{X},\mathbf{Y})=\int p(\mathbf{y^*}|\mathbf{x^*},\nu)p(\nu|\mathbf{X,Y})d\nu,$$
where the posterior distribution of $p(\nu|\mathbf{X,Y})$ can be approximated by a variational distribution of $p(\nu)$ that involves a Gaussian mixture modeling of the weight matrices across the nodes (neurons) for each layer and a simple Gaussian approximation distribution for the bias vector $\mathbf{b}$. After the dropout procedure is applied to the weight matrices repeatedly, the empirical mean and standard deviation of the predicted distribution can be used to construct a prediction interval. This approach is suitable for any models with minimal changes that does not sacrifice computational complexity or test accuracy.

\section{Simulation Studies}
\label{s:sim}

In the standard DL  architecture with fully connected nodes, recall that the causal DAG interpretation would imply that the input variables would have been randomized, independent of each other, and directly affecting the output variable, so that performance of the DL algorithm could be poor when the bias due to an imbalance caused by a latent relationship between the input and output variables is not adjusted.

\begin{figure}
    \centering
    \includegraphics[width = 0.8\linewidth,keepaspectratio]{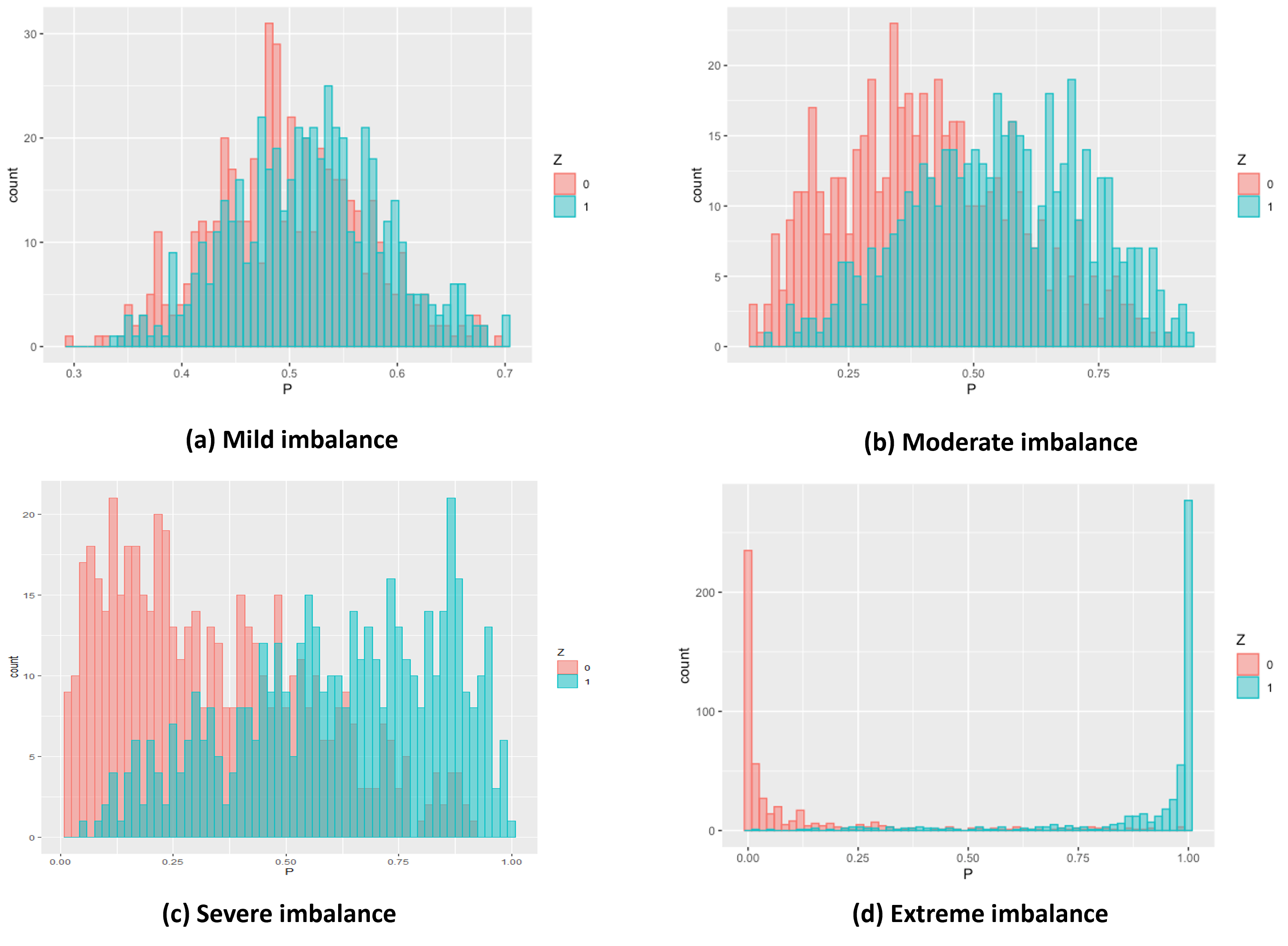}
    \caption{Degrees of imbalance level.}
    \label{fig:imbalance}
\end{figure}

To test this conjecture, first we have generated time-to-event data with or without right censoring under the true DL architecture where a linear combination of the input variables latently affects treatment assignment as in Figure \ref{fig:dag}(a). Suppose we have two intervention groups, control ($Z=0$) and treatment ($Z=1$). As in Hu et al. (2021), one thousand (1,000) observations of three continuous covariates $X_1$, $X_2$, and $X_3$, were generated from a standard normal distribution and a binary treatment indicator $Z$ was generated from a Bernoulli distribution with a probability of success of $\exp(\beta'\mathbf{X}) / \{1 + \exp(\beta'\mathbf{X})\}$, where $\mathbf{X}'=(X_1,X_2,X_3)$ and $\beta$ is a vector of coefficients that controls the degree of imbalance level in the distribution of the input variables. The probability of being assigned to the treatment (propensity score), $P$, was estimated from the logistic regression with $Z$ being the response variable and $X_1$, $X_2$, and $X_3$ as the covariates, and the ATE weights were calculated as $w=Z/P+(1-Z)/(1-P)$. Figure \ref{fig:imbalance} shows realization of the degrees of imbalance level in terms of the overlapping propensity score distributions corresponding to the coefficient vectors of $\beta'=(0.3, -0.2, 0.1)$ for mild level of imbalance, (0.75, -0.5, 0.25) for moderate level, (1.2, -0.8, 0.4) for severe level, and (6, -4, 2) for extreme level. Then the true counterfactual event times were generated from a Weibull distribution with $Y(0) =\lambda_0\{-\log(u)/\exp(h_0)\}^{1/\alpha}$ and $Y(1) = \lambda_1 \{-\log(u) / \exp(h_1)\}^{1/\alpha}$, where $\lambda_0=18$, $\lambda_1=20$, $\alpha = 2$, $u \sim \mbox{Unif}(0,1)$, $h_0 = 0.2X_1 + 0.7X_2 + 0.4X_3$, and $h_1 = -0.5X_1 - 2X_2 -0.25X_3$. The observable event times were determined by $Y = ZY(1) + (1-Z)Y(0)$ and finally the observed event times under right censoring was determined by $T=\min(Y,C)$ with an associated event indicator $I(Y<C)$ where the independent censoring variable $C$ follows a uniform distribution between 0 and $c$, which controls the censoring proportion. We have considered another scenario where there is a collider as in Figure \ref{fig:dag}(b) by assuming $X_2$ is affected by both $X_1$ and $X_3$ by a linear fashion, i.e $5X_1+5X_3$.

\begin{figure}
    \centering
    \includegraphics[width = 0.8\linewidth,keepaspectratio]{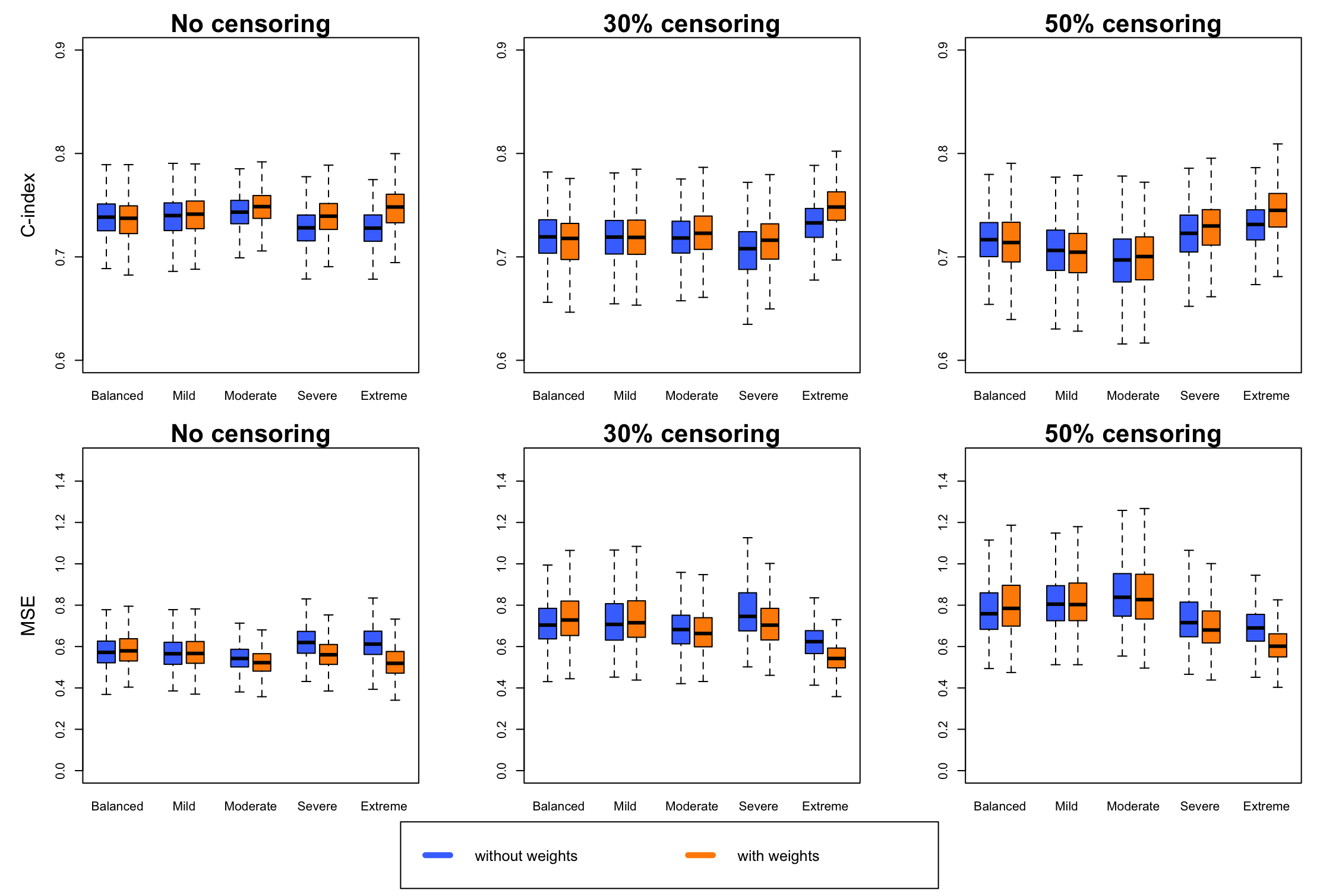}
    \caption{The C-index and MSE comparison between without and with the ATE weights in DL loss function without a collider among the input variables.}
    \label{fig:results_nocollider}
\end{figure}

In our simulation studies here, we focused on the ATE weights in our loss functions with no censoring, 30\%, and 50\% right censoring cases for a sample size of 1000, and the means of the MSE and C-index values from the 1000 iterations were used to evaluate the accuracy improvement attributed to the ATE weight functions in predicting causal individual event times. Each simulated data set was split by 70\% vs. 30\% to training and test sets, respectively, and the dropout method was applied to the training data set to avoid overfitting. Figure \ref{fig:results_nocollider} indicates that when the treatment is independent of the linear combination of the other covariates (``Balanced''), the inverse propensity score weights in our loss functions rarely influence the performance of the DL  algorithm in terms of the MSE and C-index. As the degree of imbalance level increases, however, the results indicate that the loss functions with the weight functions provide more accurate predicted values and higher prediction accuracy for causal interpretation.

\begin{figure}
    \centering
    \includegraphics[width = 0.7\linewidth,keepaspectratio]{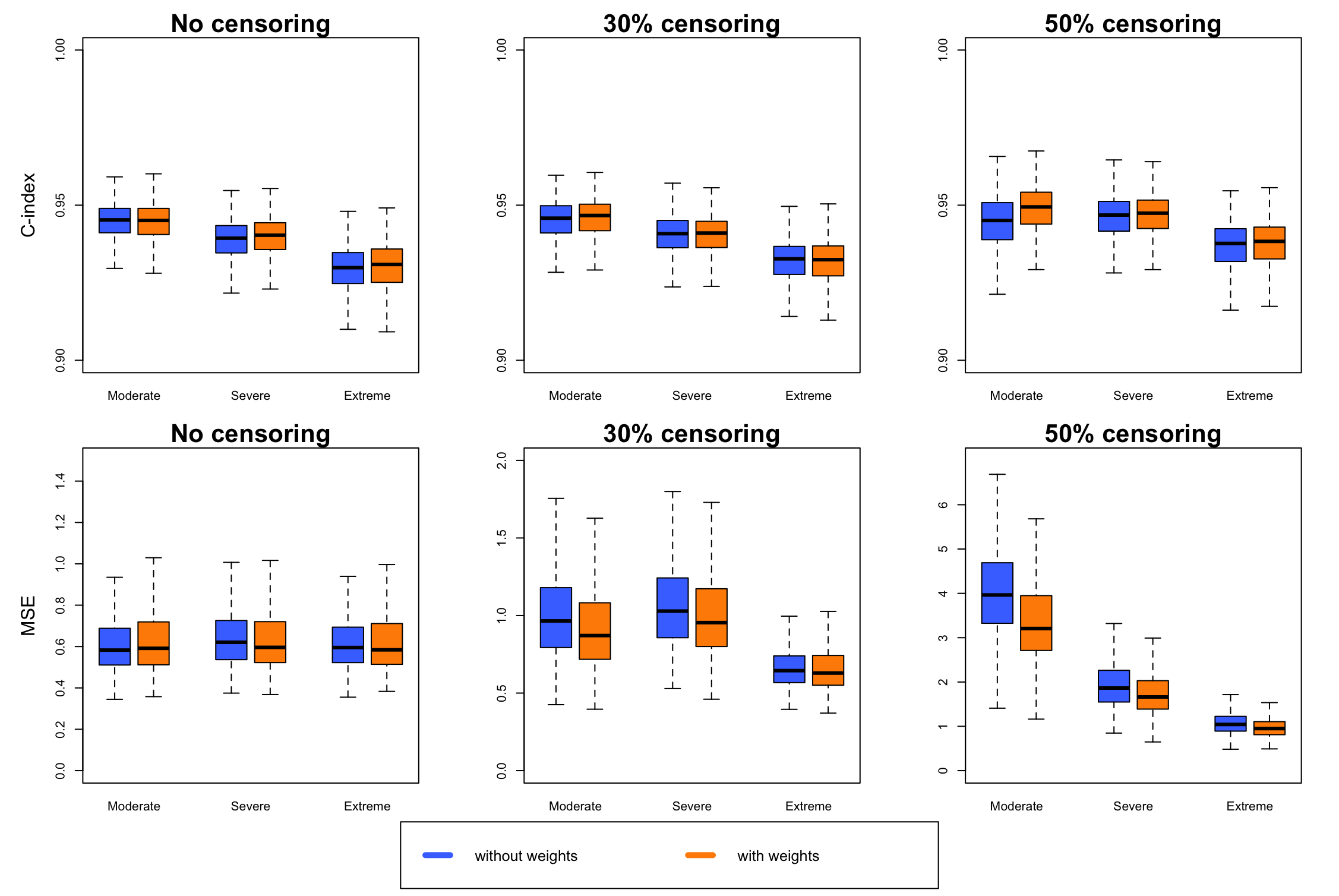}
    \caption{The C-index and MSE comparison between without and with the ATE weights in DL loss function with a collider among the input variables.}
    \label{fig:results_collider}
\end{figure}

Figure \ref{fig:results_collider} shows the simulation results for the collider case only for 3 scenarios, i.e., moderate, severe, and extreme, because the results were almost identical for the balanced and mild cases and hence to focus on the cases with noticeable differences. Larger differences can be noticed for the moderate imbalance cases with heavier censoring both in the MSE and C-index. It can be also noticed that the variations in MSE tend to decrease as both censoring proportion and imbalance level increase. This could be partially explained such that larger true values would tend to be censored, reducing the scale of the event times particularly for the calculation of the MSE, and the effects of the weights might be diluted due to the larger proportion of the censored observations under this scenario.

\section{Real Data Application}
\label{s:data}
In this section, we evaluate the performance of the proposed causal DL  algorithm in a publicly available real data set by comparing the $C$-index and MSE between with and without the ATE weights. The data set is the Molecular Taxonomy of Breast Cancer International Consortium (METABRIC) data set \citep{curtis2012genomic}. The METABRIC data set consists of 1,981 breast cancer patients' gene expression data and 21 clinical features. As mentioned in the Introduction Section, we included 4 genes via their expression levels (MKI67, EGFR, PGR, and ERBB2) with 10 clinical features (ER status, HER2 status, PR status, tumor size (Size), age at diagnosis, number of positive lymph nodes (\#LN), menopausal status (Mense), Nottingham prognostic index (NTI), hormonal therapy status (HT), and radiotherapy status (RT)). This list certainly implies potential association among the input variables, e.g., between HER2 status and ERBB2 level, PR status and PGR level, ERBB2 level and tumor size, HT status and PGR and ERBB2 levels, menopausal status and ERBB2 level, tumor size and ERBB2 level, age and ERBB2 level, chemotherapy and menopausal status, and so on. The outcome variable is time to death, possible censored, and the treatment group variable was chemotherapy. We restricted to a subset of 1,884 patients with complete data on the included covariates, where the censoring proportion is 42.14\%.

\begin{figure}
    \centering
    \includegraphics[width = 0.6\linewidth,keepaspectratio]{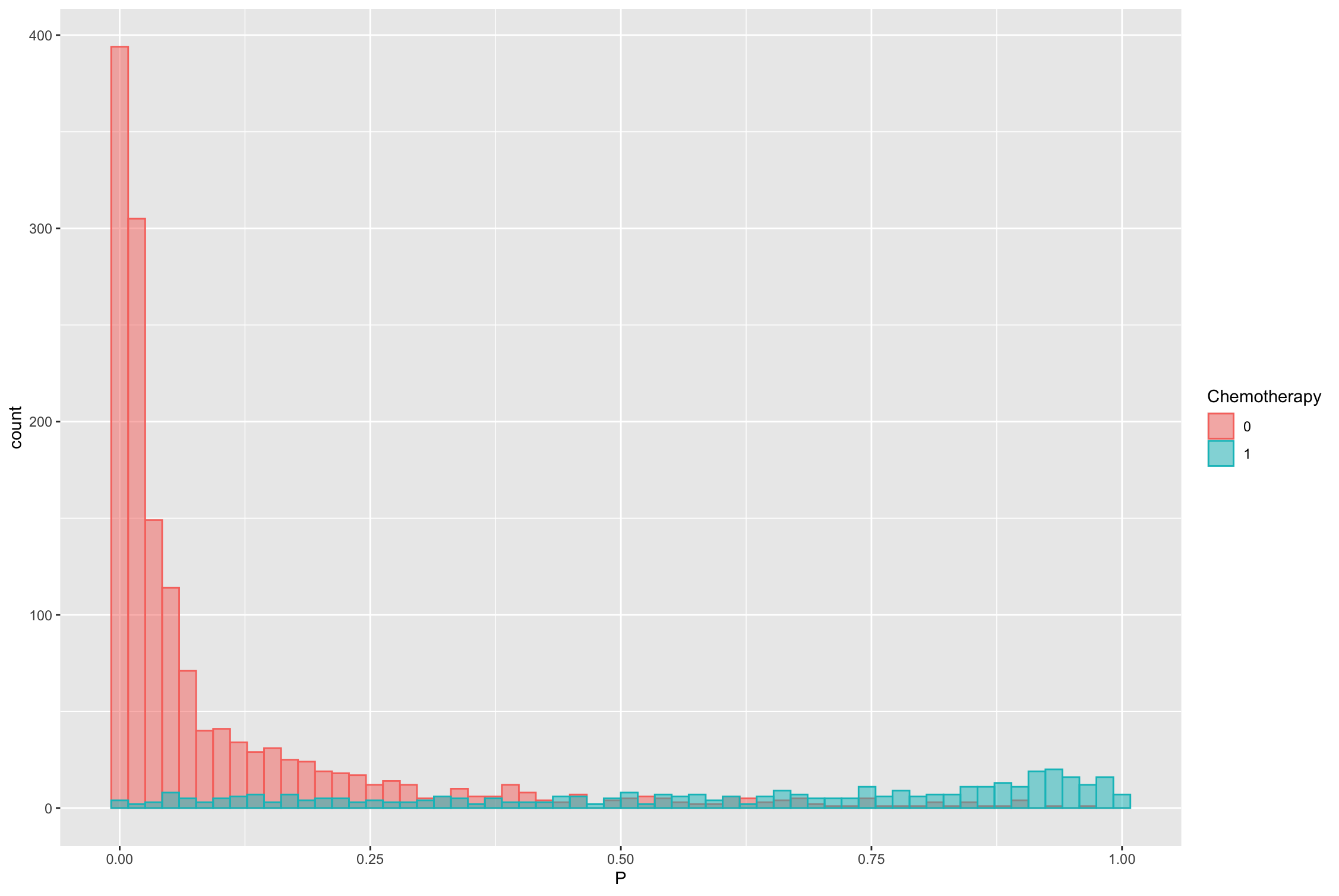}
    \caption{The overlapping histogram of estimated propensity scores from the METABRIC data set.}
    \label{fig:METABRIC}
\end{figure}

The data set was randomly splitted into 70\% training and 30\% test sets and repeated for 50 times to calculate the average performance based on the $C$-index and MSE. Note that we did the hyperparameter tuning in each of the 50 training sets separately. Figure \ref{fig:METABRIC} shows the overlapping histogram of the estimated propensity scores from the logistic regression, indicating that a substantial proportion of the control patients could contribute their counterfactual information toward the treatment group.

The results (Table ~\ref{tab:METABRIC_results}) showed that the C-index values (SD) were 0.583 (0.018) (95\%CI=[0.578, 0.588]) and 0.594 (0.020) (95\%CI=[0.588, 0.600]), and the MSE values (SD) were 1.042 (0.087) (95\%CI=[1.018, 1.066]) and 0.889 (0.088) (95\%CI=[0.865, 0.913]) without and with the ATE weights, respectively. Non-overlapping 95\% CIs indicate statistical significance among the comparisons between without and with the weights in the loss function.

\begin{table}[]
\centering
\caption{C-index and MSE (mean (SD)) with and without the ATE weights in DL loss function for METABRIC data set}
\vspace{.3cm}
\label{tab:METABRIC_results}
\begin{tabular}{c|c|c}
\hline
        & without $w$     & with $w$        \\
\hline
C-index & 0.583 (0.018) & 0.594 (0.020) \\
MSE     & 1.042 (0.087) & 0.889 (0.088) \\ \hline
\end{tabular}
\end{table}

The trained and validated DL  algorithm can be used to predict causal individual event times with or without a chemotherapy. We have selected two patients with most similar input features from the observed data set, one with chemotherapy and the other without it (Table ~\ref{tab:prediction}). After 100 runs of the trained model, as typically done in the literature, the means of the predicted posterior distributions were 15.72 months (95\% prediction interval = [13.35,18.50]) and 77.32 months (95\% prediction interval = [54.80,109.09]) for Pt. \#1 and Pt. \#2, respectively. Their observed event times were 14.4 and 75.7, respectively. The {\em counterfactual} predicted time would have been 58.32 months (95\% prediction interval = [41.34, 82.26]) if Pt. \#1 would not have taken the chemotherapy and would have been 39.60 months (95\% prediction interval = [27.43, 57.16]) if Pt. \#2 would have taken the chemotherapy. In the original data set, the observed event times for the patients that had the same binary covariate values and similar continuous covariate values to Pt. \#1 without chemotherapy and Pt. \#2 with chemotherapy were 52.97 months and 36.4 months, respectively, which shows that the predictions seem reasonable.

\begin{figure}
    \centering
    \includegraphics[width = 0.9\linewidth,keepaspectratio]{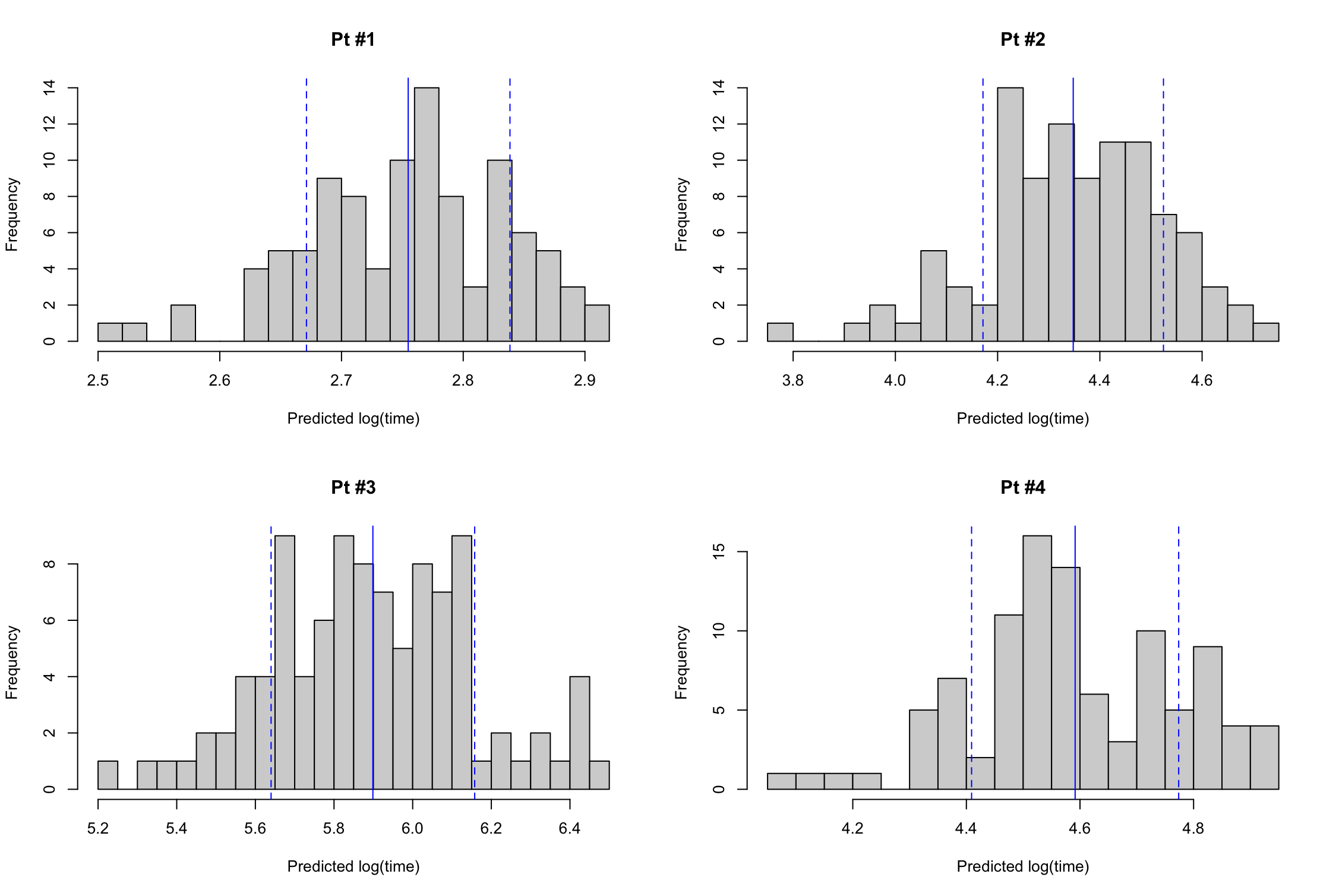}
    \caption{Histogram of predicted values for four selected patients (solid blue line indicates the mean and dashed blue lines indicate mean +/- SD) }
    \label{fig:METABRIC_histogram}
\end{figure}

Now we have considered arbitrary input variable vectors that were not in the observed data; a ER-negative, HER2-negative, PR-positive, and postmenopausal subject with the mean values of tumor size, age at diagnosis, the number of positive lymph nodes, Nottingham prognostic index, and 4 specified gene expression levels, who also received a hormonal therapy and a radiation therapy. For these extrapolated cases, a predicted individual event time to death would be 345.8 months (95\% prediction interval = [220.1, 604.1]) if she would have taken the chemotherapy (Pt. \#3), and it would be 94.5 months (95\% prediction interval = [69.1, 140.7]) otherwise (Pt. \#4). Figure \ref{fig:METABRIC_histogram} shows the histograms of the predicted posterior distributions on a log scale for all those 4 cases considered here for prediction, which visually shows that prediction of the outcome variable from extrapolated input variable vectors using the DL  algorithm could result in more serious uncertainty, with wider prediction intervals. More specifically, there were fewer subjects in the training data set who had similar covariate values to Pt. \#3 than to Pt. \#4, which led to a much wider prediction interval for Pt. \#3.

\begin{table}[]
\centering
\caption{Selected patients from the observed data for prediction}
\vspace{.3cm}
\label{tab:prediction}
\begin{adjustbox}{width=\textwidth}
\begin{tabular}{c|ccccccccccccccc}
\hline
Pt. \#&Chemo &	ER &HER2&PR&Size&Age& \#LN & NTI &HT&	Mense& RT&	MKI67&	EGFR&	PGR&	ERBB2 \\  \hline
1& 1&	0&	0&	0&	25&	45.74&	8&	6.05&	0&	0&	1&	5.96&	6.19&	11.92&	5.27 \\
2& 0&	0&	0&	0&	25&	60.51&	0&	3.05&	0&	1&	1&	6.27&	7.11&	9.93&	5.63\\
\hline
\end{tabular}
\end{adjustbox}
\end{table}

\section{Discussion}
\label{s:discuss}

 Machine learning algorithms including DL  have been popular prediction tools, but without causal interpretation the predicted results even from a validated model might not be adequate to be applied to a future patient for intervention due to potential bias introduced by the confounding factors. In this study, the DL  architecture was modeled as a causal DAG, and a novel DL  algorithm with a loss function inherently adjusting for the potential bias due to a latent causal structure among the input variables was developed to estimate causal individual event times with or without right censoring. The proposed causal DL  algorithms provided the individual event time estimates closer to the observed values when there is a causal structure among the input variables that causes imbalance between intervention groups. Therefore we strongly recommend that the DL  algorithm be weighted accordingly for causal interpretation of the predicted values for future intervention. The proposed approach does not need estimation of the causal DAG from the observed data, which often requires strong unidentifiable assumptions. The logistic regression model was adopted in this paper to estimate the propensity scores, which is known to be simple yet robust, but other approaches such as calibration weighting \citep{chan2015} can be also used. To predict a future observation, however, like all the other machine learning algorithms, care is needed against excessive extrapolation.

 Further investigation into estimation of the average treatment effects through the DL  algorithm accounting for both sampling distributions and uncertainty of the model uncertainty and further extension to competing risks data might merit future research. Analytic investigation into performance of the weighted DL  algorithm might also merit future research.

\section*{Acknowledgments}
This research was supported in part by the University of Pittsburgh Center for Research Computing through the resources provided.


\bibliographystyle{biom}
\bibliography{mybib}

\end{document}